\documentclass[prb,twocolumn,showpacs,groupedaddress,preprintnumbers]{revtex4}
\usepackage{amssymb}
\usepackage{epstopdf}
\usepackage{graphicx}
\usepackage{dcolumn}
\usepackage{amsmath}
\usepackage{amsfonts}

\begin{document}

\title[Short Title]{Ultrafast vibrational motion of carbon nanotubes in different pH environments}

\author{Kotaro Makino$^{1}$}
\author{Atsushi Hirano$^{1}$}
\author{Kentaro Shiraki$^{1}$}
\author{Yutaka Maeda$^{2,3}$}
\author{Muneaki Hase$^{1,3}$}
\email{mhase@bk.tsukuba.ac.jp}
\affiliation{$^{1}$Institute of Applied Physics, University of Tsukuba, 1-1-1 Tennodai, Tsukuba 305-8573, Japan}
\affiliation{$^{2}$Department of Chemistry, Tokyo Gakugei University, 4-1-1 Nukuikitamachi, Koganei, Tokyo 184-8501, Japan}
\affiliation{$^{3}$PRESTO-JST, 4-1-8 Honcho, Kawaguchi, Saitama 332-0012, Japan}


\begin{abstract}
  We have used a femtosecond pump-probe impulsive Raman technique to 
  explore the ultrafast dynamics of micelle suspended single walled carbon nanotubes 
  (SWNTs) in various pH environments. The structures of coherent phonon 
  spectra of the radial breathing modes (RBMs) exhibit significant pH 
  dependence, to which we attribute the effect of the protonation at the surface 
  of SWNTs, resulting in the modification of electronic properties of 
  semiconductor SWNTs. Analysis of the time-domain data using a time-frequency transformation 
  uncovers also a second transient longitudinal breathing mode, which vanishes after 1 ps of the 
  photoexcitation. 
\end{abstract}

\pacs{78.47.J-, 78.67.Ch, 63.22.Gh, 63.20.kd}

\maketitle
Carbon nanotubes (CNTs) are one of the most intriguing nano-materials, which 
are useful for molecular devices, such as a chemical sensor\cite{Modi03} and a nano-machine 
capable for medical science.\cite{Zhang08} One potential way to realize the nano-machine by the 
use of CNTs, which can work in biological systems, e.g., a drug delivery system,  
is to use them in solution with proteins.\cite{Zhang08} 
Therefore, CNT devices will be used in various pH environments in biological systems, 
e.g., the typical pH of human arterial blood is $\approx$ 7.4, that is weakly alkaline. 
Since the chemical reactivity of CNTs is dominated by the fundamental physical processes 
on their surface, investigation of dynamics at surfaces, such as charge transfer,\cite{Rao97} 
exciton-plasmon coupling,\cite{Bondarev09}  and electron-phonon energy transfer,\cite{Hertel00} 
is crucial to prepare suitable environments for CNT devices. 

Single-walled carbon nanotubes (SWNTs) possess simplest structure of CNTs, 
and it has metallic or semiconducting conductivity, depending on the arrangement of 
carbon atoms, referred as roll-up vectors.\cite{Saito03}
Recently frequency-domain spectroscopy has revealed the diameter-selective 
Raman scattering from radial breathing modes (RBMs) in SWNT,\cite{Jorio01,Dresselhaus05} 
where the frequency of the RBMs depended on the inverse of the diameter.\cite{Rao97}
Moreover, 
Strano {\it et al.} have investigated the change in the Raman spectra of RBM induced 
by the protonation and showed that the protons change electronic structure of SWNT, i.e., 
protonation is band-gap-selective and consequently the condition of resonant Raman 
scattering was modified, depending on the pH values.\cite{Strano03}
Coherent phonon spectroscopy is another potential method to study phonon 
dynamics in time-domain. 
Up to date, there have been only several studies on ultrafast phenomena in 
CNTs,\cite{Hertel05,Lim06,Gambetta06,Kato08,Kim09} and, early time 
stage dynamics of photo-excitation of electronic system and subsequent phonon 
excitation and relaxation was not been well explored. Motivated by the observation of 
the early time stage dynamics of electron-phonon coupling, L\"{u}er {\it et al}. have 
recently investigated the electron-phonon coupling by means of coherent phonon 
spectroscopy with 10 fs time-resolution,\cite{Luer09} where they focused their attention to the 
high frequency optical modes rather than the RBMs. 

In this paper, the coherent oscillation of the RBMs in SWNTs has been measured 
in various pH environments to show that 
the subpicosecond relaxation dynamics as well as the coherent phonon spectra of 
the RBM are very sensitive to the change in the electronic structure of SWNT upon the 
protonation. 

The samples of SWNTs were produced via the high-pressure catalytic CO decomposition 
(HiPco) process.  To disperse SWNTs in aqueous solution, we used sodium dodecyl sulfate (SDS).  
Solution containing 100 mM SDS in distilled water was mixed with SWNT powder. 
The SWNTs were dispersed by ultrasonication for 30 min. at 20$ ^{\circ } $ C. 
Non-dispersed SWNTs were removed by filtration of the solution using Whatman filter paper No.41. 
Successively, they were diluted 3-fold into 50 mM citrate-phosphate-borate 
buffer solutions at different pHs (3, 4, 5, 6 and 7), and they were filled in the 5-mm thick quote cells. 
Absorption spectra obtained for SWNTs in SDS solution (Fig. 1) showed that the sharp 
peaks corresponding to the exciton bands\cite{Wang05} appeared on the background absorption, 
which was assigned to $\pi$-plasmon contribution due to graphitic carbon or nanotube bundles.\cite{Landi05}  
As shown in the inset of Fig. 1, 
the absorbance at 820 nm due to exciton resonance gradually increased as the pH value increased. 

Time-resolved transient transmission of the sample was measured by employing a fast scanning delay 
technique using a mode-locked Ti:sapphire laser. Nearly collinear, pump and 
probe pulses (20 fs pulse duration; 850 nm (=1.46 eV) wavelength; 80 MHz repetition 
rate) were focused to a 70 $ \mu $m spot into the sample cell. The average power of the 
pump and probe beams were fixed at 40 and 3 mW, respectively. 
The transient transmission ($\Delta T/T$) was recorded as a function of the  
time delay $\tau$ between the pump and probe pulses. 
\begin{figure}[htbp]
\includegraphics[width=80mm]{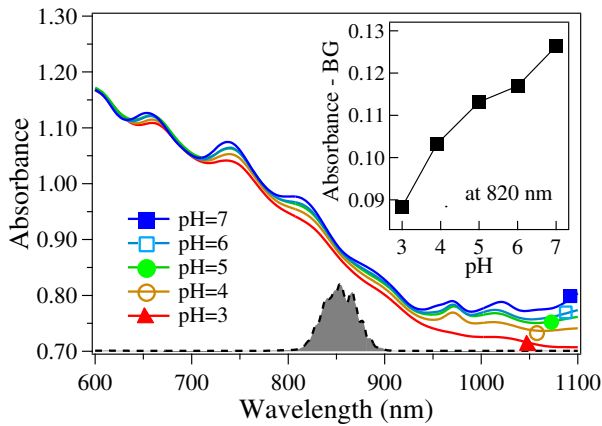}
\caption{(Color online) Absorption spectra obtained in SWNT/SDS samples at various pH values. 
The dashed line with gray shading represents laser spectrum centered at 850 nm, showing full 
bandwidth of $\sim$ 100 nm. The inset shows the absorbance at 820 nm as the function of pH 
value obtained by subtracting background (BG) absorption based on $\pi$-plasmon model 
(see the text). 
}
\label{fig1}
\end{figure}

\begin{figure}[htbp]
\includegraphics[width=80mm]{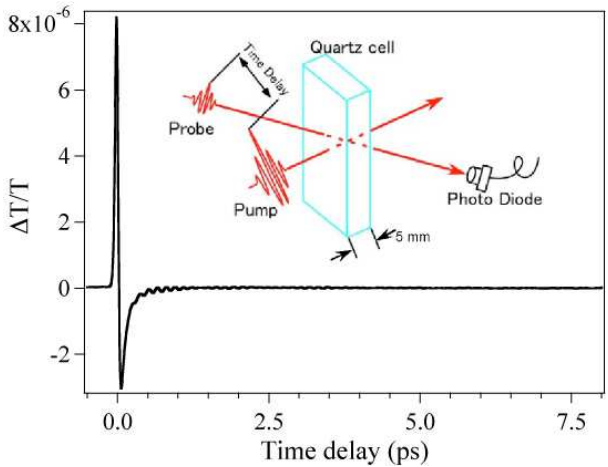}
\caption{(Color online) Time-resolved transmission for SWNT/SDS sample at pH = 3. Coherent 
RBMs are observed just after the transient electronic response at $\tau$ = 0 ps. 
Pump-probe setup around the quart cell is schematically shown in the inset. }
\label{fig2}
\end{figure}
Figure 2 shows $\Delta T/T$ signal observed for SDS 
suspended SWNTs in solution at pH = 3. The transient electronic response appears 
at $\tau$ $\approx$ 0 ps and it is due to the change in the third-order nonlinear susceptibility 
$\chi^{(3)}$ from water solution\cite{Lim06} as well as due to the excitation and subsequent relaxation 
of photoexcited carriers in SWNTs. The transient electronic response is followed by the coherent 
oscillations due to the RBMs, whose frequency $\omega_{RBM}$ inversely 
depends on the diameter of SWNT by,\cite{Rao97}  
\begin{equation}
\omega_{RBM} = 7.44 [THz] / R_{t} [nm],
\label{eq1}
\end{equation}
where $R_{t}$ is the diameter of SWNT. 
By subtracting the carrier responses from the time-domain data in Fig. 2,\cite{Note1} only the oscillatory 
components due to the coherent RBMs are obtained as shown in Fig. 3. 
The $\Delta T/T$ signal, thus obtained in the different pH environments (pH = 3, 4, 5, 6, and 7),  
indicate that the amplitude of the coherent RBMs seems to increase with increasing the pH value, 
implying that the appearance of the RBM is sensitive to the pH environment. 
The coherent RBMs are fit by the linear combination of the damped harmonic oscillations, 
\begin{eqnarray}
\frac{\Delta T}{T} = A_{L}e^{- t/\tau_{L}} \cos[2\pi (\nu_{L} + \alpha_{L}t)t + \varphi_{L}]\nonumber\\
+ A_{H}e^{- t/\tau_{H}} \cos[2\pi (\nu_{H} + \alpha_{H}t)t + \varphi_{H}] ,
\label{Eq3}
\end{eqnarray}
where $A_{L}$ ($A_{H}$), $\nu_{L}$ ($\nu_{H}$), $\tau_{L}$ ($\tau_{H}$),  $\alpha_{L}$ ($\alpha_{H}$), 
and $\varphi_{L}$ ($\varphi_{H}$) are the amplitude, the {\it static} frequency, the 
dephasing time, the frequency chirp, and the initial phase of the dominant coherent RBMs of the 
lower ($\nu_{L}$ $\approx$ 6.4 THz) and the higher ($\nu_{H} $ $\approx$ 7.2 THz) 
modes as shown in the Fourier transformed (FT) spectra in Fig. 4. 
Note that satellite peaks appear at the lower frequency side of the 6.4 THz mode (Fig. 4), implying 
inhomogeneity of the tube diameters.\cite{Dresselhaus05}  

By using the Kataura's plot it is possible to extract which chirality (n, m) of SWNTs satisfies resonant 
condition for the Raman scattering and to identify the electronic state of SWNT, as semiconducting or 
metal.\cite{Kataura99}  
In our sample the diameter of SWNT has a range of 0.9 - 1.3 nm. 
Therefore, both the dominant lower RBM at 6.4 THz and the higher one at 7.2 THz are considered to 
be the RBMs from the semiconductor SWNTs. In fact, according to Eq. (1), the peak of 6.4 THz 
corresponds to $\approx$ 1.16 nm tubes, while the peak of 7.2 THz corresponds to $\approx$ 
1.03 nm tubes, where the chirality of the lower mode can be (13, 2) and that of the higher mode 
can be (12, 1).\cite{Goupalov06} 
Considering the photon 
energy used in the present study (850 nm), we can conclude that semiconducting SWNTs are 
excited through resonant impulsive stimulated Raman process.\cite{Stevens02} 
Because of the broad bandwidth ($\sim$ 100 nm in full) of the laser used, the excitonic 
transitions around the 
critical point of E$_{22}$ \cite{Saito03} are available at 800 - 900 nm.  
Note that our laser can excite also transition corresponding to E$_{11}$ critical point 
beyond the gap. In this case, however, the intensity of the RBM would be negligibly small 
compared to those excited through the more resonant E$_{22}$ band. 

\begin{table*}
  \caption{Main parameters obtained from the fittings of the oscillatory traces in Fig. 3 by using 
  Eq. (2). }
\begin{ruledtabular}
  \begin{tabular}{ccccccc}
     pH & $\nu_{L}$ (THz) & $\nu_{H}$  (THz) & $\tau_{L}$ (ps) & $\tau_{H}$ (ps) & $\alpha_{L}$ (ps$^{-2}$) & $\alpha_{H}$ (ps$^{-2}$)\\
    \hline
    3 & 6.34 $\pm$ 0.02 & 7.18 $\pm$ 0.02 & 0.80 $\pm$ 0.1 & 1.97$ \pm$ 0.1 & 0.002 $ \pm$ 0.01& -0.015 $\pm$ 0.005\\
    4 & 6.35 $\pm$ 0.02 & 7.19 $\pm$ 0.02 & 0.87 $\pm$ 0.1 & 1.95$ \pm$ 0.1 & 0.0002 $ \pm$ 0.01& -0.018 $\pm$ 0.005\\
    5 & 6.33 $\pm$ 0.02 & 7.21 $\pm$ 0.02 & 0.85 $\pm$ 0.1 & 2.02$ \pm$ 0.1 & 0.002 $ \pm$ 0.01& -0.024 $\pm$ 0.005\\
    6 & 6.34 $\pm$ 0.02 & 7.22 $\pm$ 0.02 & 0.94 $\pm$ 0.1 & 2.10$ \pm$ 0.1 & 0.002 $ \pm$ 0.01& -0.027 $\pm$ 0.005\\
    7 & 6.32 $\pm$ 0.02 & 7.21 $\pm$ 0.02 & 1.08 $\pm$ 0.1 & 1.99$ \pm$ 0.1 & 0.006 $ \pm$ 0.01& -0.024 $\pm$ 0.005\\

  \end{tabular}
  \end{ruledtabular}
\end{table*}

\begin{figure}[htbp]
\includegraphics[width=80mm]{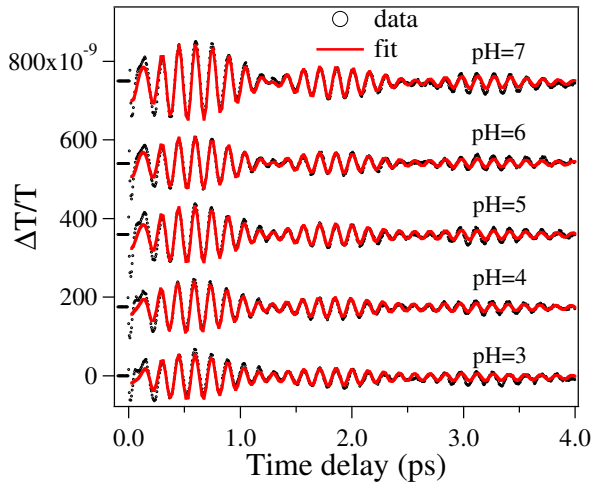}
\caption{(Color online) Time domain signal from the coherent RBM obtained by subtracting carrier 
responses from the time-domain transmission at various pH values. The solid lines represent 
the fit to the time-domain data with Eq (2).} 
\label{fig3}
\end{figure}
We find in Fig. 3 that Eq. (2) fits the time-domain data well, although there is 
a significant discrepancy between the data and the fit at the early time stage, i.e., within the first oscillation 
cycle at $\leq$ 300 fs. 
From the obtained fitting parameters (Table 1) we notice that the {\it static} frequency of the coherent RBMs 
does not depend on the pH value, i.e., 
$ \nu_{L}$ = 6.33 $\pm$ 0.02 THz, $ \nu_{H} $ = 7.20 $\pm$ 0.02 THz, indicating 
that the {\it static} diameter of SWNTs is not effectively changed when the pH value is varied.
On the other hand, the frequency chirp of the higher mode $\alpha_{H}$ is always negative, 
slightly depending on the pH value. This fact suggests that the {\it dynamic} frequency of the higher RBM 
is dependent on the time delay; the negative chirp means that the transient frequency is blue-shifted 
at early time delays, followed by the red-shift. Such the blue-shift might be explained by the electron-phonon 
decoupling induced by the carrier doping, as observed for the E$_{2g2}$ mode in graphite,\cite{Ishioka08} 
and for the RBM in metallic SWNTs.\cite{Farhat09} 
Moreover, the dephasing time of the lower mode significantly increases with increasing the pH value; 
$ \tau_{L} $ changes from 0.80 to 1.08 ps, while $\tau_{H}$ is almost unchanged as shown in Table 1. 
Note that we cannot find any systematic change in the initial phase of the dominant coherent RBMs 
for the different pH values (not shown). 

\begin{figure}[htbp]
\includegraphics[width=80mm]{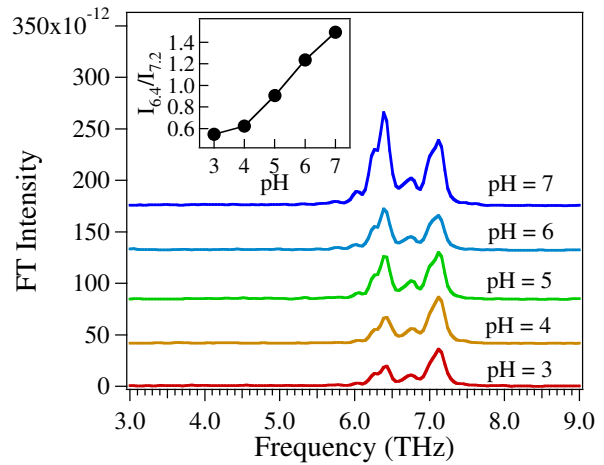}
\caption{(Color online) The Fourier transformed spectra obtained from the time domain signal for 
various pH values. 
The two dominant peaks appear at 6.4 THz (lower mode) and at 7.2 THz (higher mode). 
The inset represents the ratio of the intensity of the lower mode to the higher mode, indicating the lower mode is 
enhanced in the order of pH = 3, 4, 5, 6, and 7. }
\label{fig4}
\end{figure}
Figure 4 shows coherent phonon spectra obtained from the traces in Fig. 3 by using FT. It is 
found that the lower RBM at 6.4 THz is strongly 
enhanced as the pH value increases, indicating that the lower protonation, the stronger intensity of the RBM 
at 6.4 THz. 
The effect of the protonation (and consequently the effect of decreasing the pH value) on the 
RBMs was in past examined by using Raman scattering, where the reduction of the Raman 
intensity of the semiconducting SWNTs was observed with increasing acidification (deprotonation), and the 
authors claimed that the acidification of the solution of surfactant-dispersed SWNT in water 
resulted in a reaction with protons at the sidewall of CNT.\cite{Strano03} 
Here, the protonation can be considered to capture free electrons from semiconducting SWNTs. 
Based on these arguments, H$^{+}$ 
ions would act as acceptors, changing the electronic property of semiconducting SWNT from $p$-type 
at pH = 3 into nearly undoped at pH = 7.\cite{Ostojic04} 
This may explain why the absorbance gradually increases as the pH value increases (Fig. 1) 
because of the reduced oscillator strength due to a partly emptied valence band.\cite{Kim08} 
Note that the factor of the increase in the absorbance due to the exciton resonance ($\approx$1.43 at 820 nm 
as shown in the inset of Fig. 1) 
is smaller than the experimental observation ($ \sim $4.0 for the lower RBM as in Fig. 4), 
implying the existence of other contributions to the observed enhancement of the RBM amplitude in Fig. 4.

From Table 1 we found (i) 
the dephasing time of the lower RBM depends on the pH value, while that of the higher RBM 
at 7.2 THz is not dependent on pH, and (ii) the dephasing time of the coherent RBM at 6.4 THz is 
shorter than that of the higher RBM at 7.2 THz. 
Regarding to (i) above, the pH dependence of the depahsing time would be related to the 
carrier-phonon interaction at early time delays. 
In fact, Ostojic {\it et al.} reported that the early stage relaxation process of photoexcited carriers 
in SWNTs was found in similar time region, 0.3 - 1.2 ps, and the relaxation time depended on pH 
values; the lower pH value, the faster relaxation time (destroy the slow component).\cite{Ostojic04} 
They also argued that based on the Burstein-Moss effect\cite{Burstein54}  larger diameter tube 
(consequently lower frequency RBM) is more 
susceptible to this pH dependence because of the smaller accepter binding energies of larger 
diameter tubes. 
In addition to the effect described above, SWNT becomes $p$-type 
semiconductor as the pH value decreases.\cite{Ostojic04}  If this is the case, the lifetime of carriers is rather 
short at the lower pH values because of the shorter lifetime of the photogenerated 
holes,\cite{Ganikhanov98}  and therefore the dephasing time of the coherent RBM, which is coupled 
with the photogenerated carriers, is also shorter. These considerations are consistent with 
the behavior of $\tau_{L}$ for the different pHs in Table 1.  
Note that a possible reason why the depahsing time of the higher RBM is not dependent on pH 
is the electron-phonon decoupling, which would occur in a time scale of a few picoseconds,\cite{Ishioka08} 
and would stronger for the higher RBM.\cite{Farhat09} 

In the present study, RBMs are considered to be coupled with electrons through 
the optical deformation potential interaction,\cite{Jiang05} i.e., the matrix element $|M_{e-ph}|$ of the 
electron-phonon coupling can be described by 
$|M_{e-ph}|^{2} = |a/R_{t}^{2} + (\nu b/R_{t})cos3\theta |^{2}$, where $a$ and $b$ are the constants, 
$\nu$ [= mod($n-m$, 3)] is the chiral index, and $\theta$ is the chiral angle.\cite{Goupalov06,Yin07}  
Thus the electron-phonon coupling strength decreases if the diameter of the SWNT ($R_{t}$) increases. 
Consequently, regarding to the note (ii) above, the shorter dephasing time of the lower frequency RBM, 
corresponding to the larger 
diameter tubes, cannot be explained by the electron-phonon coupling, since the electron-phonon 
matrix element of the deformation potential interaction becomes weak when the diameter of SWNT 
increases.\cite{Jiang05,Yin07} 
Instead of considering electron-phonon coupling, inhomogeneous damping of the lower frequency RBM plausibly 
accounts for the shorter dephasing time of the lower RBM than that of the higher RBM.\cite{Note2}

From the considerations above, the reason why the FT intensity enhancement in Fig. 4 depends 
on the chirality may be explained by the increased dephasing time of the lower mode ($\tau_{L}$), that is a factor 
of $\tau_{L}$(pH=7)/$\tau_{L}$(pH=3) = 1.35 from Table 1. Since the amplitude $A$ of the coherent phonon would be 
proportional to $\gamma^{-1} (= \tau$), 
where $\gamma$ is the damping,\cite{Hase98} we expect the enhancement of the FT intensity ($A^{2}$) 
for the lower RBM by a factor of 1.82. 
This value is smaller than the experimental observation ($\sim$ 4 as in Fig. 4), however, 
the effect of the change in $\tau_{L}$ would dominantly contribute to the enhancement of 
the FT intensity. 

In order to analyze ultrafast relaxation of the coherent RBM in detail, we first utilize the continuous wavelet 
transform (CWT) method\cite{Hase03}, to obtain the time-frequency chronograms from the time-domain 
datum for pH=3 presented in Fig. 3. Note that there was no considerable dependence of the CWT spectra 
on the pH value, excepting the ratio of the spectral intensities of the coherent RBMs. 
\begin{figure}[htbp]
\includegraphics[width=80mm]{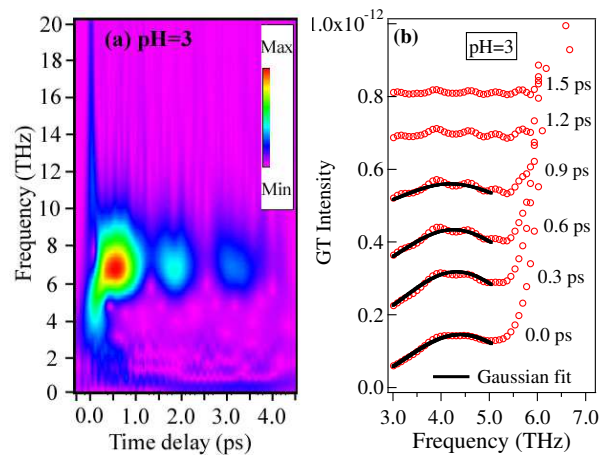}
\caption{(Color online) (a) The CWT chronogram for pH = 3. 
(b) Enlarged GT spectra around 4 THz for pH = 3, in which the broad band appears at $\approx$ 4.4 THz. 
The solid curve in (b) is the fit to the GT spectra with a Gaussian function. }
 \label{fig5}
\end{figure}
In Fig. 5(a) it is found that a coherent phonon response appears at 4 - 5 THz at the early 
time delay of near $\tau$ = 0 ps, followed by the strong peak at  6 - 8 THz due to the coherent 
RBMs. This new mode at $\approx$ 4 THz decay so fast ($\leq$ 1 ps, i.e., too broad in the frequency 
domain) that the FT analysis in Fig. 4 could not 
detect the mode. To gain more insight on the new mode at $\approx$ 4 THz, 
we use also a discrete Gabor transform (GT),\cite{Gabor46} which is expressed by the following equation,
\begin{equation}
I(\omega,\Delta\tau) =  \int_{0}^{\infty} \Bigl(\frac{\Delta T}{T}\Bigr)\exp\Bigl[-\Bigl(\frac{t-  \Delta\tau}{\sigma}\Bigr)^{2}\Bigr] \exp(-i \omega t) dt
\label{eq3}
\end{equation}
Here, $ \Delta \tau $ and $ \sigma $ (= 1.0 ps) represent the time delay and the width of the 
window function, respectively. In this method, the window function (Gaussian) extracts oscillatory components 
at $ \Delta \tau $, and thus GT method achieves time-resolved FT.\cite{Hase02} 
As shown in Fig. 5(b), it is revealed that a broad peak at $\approx$ 4.4 THz appears 
within $\approx $1 ps. This peak frequency is consistent with that of the new mode observed in CWT spectra in Fig. 5(a), 
and appears to be the longitudinal breathing (LB) mode of SWNT, that is another coherent mode identified recently 
in molecular dynamics simulations.\cite{Traian06}  This LB mode would thermalize fast due to 
inertial confinement along the tubular length.\cite{Jeschke07} 

In conclusion, we have explored the ultrafast vibrational dynamics of coherent RBMs in real time 
using the femtosecond transmission technique based on impulsive Raman scattering. 
The two dominant RBMs were observed at 6.4 and 7.2 THz, whose spectral intensities 
depended on the pH value.  The enhancement of the lower RBM (6.4 THz) at the higher pH values 
could be attributed to the change in the electronic structure of SWNT 
from $p$-doped to nondoped nature by the deprotonation effect. 
Dephasing time of the lower RBM extracted from the time-domain datum exhibited pH 
dependence, which was explained by the electron-phonon coupling through deformation potential interaction, while that of the 
higher RBM did not depend on the pH value, implying existence of the electron-phonon decoupling in a few picoseconds. 
Analysis of the time-domain data by both the continuous wavelet transform and the Gabor transform 
methods revealed the new peak at $\approx$ 4.4 THz in the transient response within 1 ps, which 
appeared to be the signature of the longitudinal breathing mode of SWNT. 

The authors acknowledge T. Dumitrica for stimulating discussions and T. Kameda for his 
helpful comments on the coherent phonon spectra. This work was partly supported by the Step-up Support Program 
for the Young Scientist from University of Tsukuba.

\end{document}